# Optical absorption and photoluminescence spectroscopy of the growth of silver nanoparticles

P. Gangopadhyay,[1,*] R. Kesavamoorthy,[1] Santanu Bera,[2] P. Magudapathy,[1] K. G. M. Nair,[1] B. K. Panigrahi,[1] and S. V. Narasimhan[2]

[1]Materials Science Division, Indira Gandhi Centre for Atomic Research, Kalpakkam-603102, India

[2]Water and Steam Chemistry Laboratory, BARC Facilities, Kalpakkam-603102, India

## Abstract

Results obtained from the optical absorption and photoluminescence (PL) spectroscopy experiments have shown the formation of excitons in the silver-exchanged glass samples. These findings are reported here for the first time. Further, we investigate the dramatic changes in the photoemission properties of the silver-exchanged glass samples as a function of postannealing temperature. Observed changes are thought to be due to the structural rearrangements of silver and oxygen bonding during the heat treatments of the glass matrix. In fact, photoelectron spectroscopy does reveal these chemical transformations of silver-exchanged soda glass samples caused by the thermal effects of annealing in a high vacuum atmosphere. An important correlation between temperature-induced changes of the PL intensity and thermal growth of the silver nanoparticles has been established in this Letter through precise spectroscopic studies.

[*] Electronic address: pganguly@igcar.ernet.in

Quantum-size behavior of nanomaterials has opened new directions in many fields [1,2] of current research and modern technologies, such as, electronic logic operations [3], catalysis reactions [4], nanotweezers [5], etc. A great deal of effort also has been put to develop and characterize the nanoscale materials by various means. Among different methods of sample preparations [6,7], silver for sodium in soda glasses through the process of ion-exchange has prompted considerable research activities during the last few years because of its' immense potential applications in optoelectronic and optical switching devices [8]. In most of the related studies, characteristic surface-plasmon-resonance (SPR) of silver nanoparticles have always been given the maximum importance due to the fast optical response (~few psec) of nanoscale metal particles in the visible range of light [9,10]. In spite of the technological significance, precise spectroscopic studies to characterize the thermal stability of these optical nanomaterials are still only successful to a limited extent [11-13].

Thermal stability of optical materials of this nature is an important issue and hence detailed optical and chemical studies of the materials as well after various thermal processing is of great interest from fundamental point of view and for technological importance. In this letter, we plan to elucidate the thermal stability and the corresponding changes in optical and chemical properties of these nanocomposite samples as a function of postannealing temperature. Further, we enucleate the origin of PL in the silver-exchanged glass samples, to best of our knowledge, is not understood clearly in the literature [11,14]. Depending on the thermal history, we report that silver could be in different oxidation states in the ion-exchanged soda glass samples and this charge state of silver play a vital role in dictating the optical properties of these nanomaterials. Optical absorption, photoluminescence and x-ray photoelectron spectroscopy (XPS) techniques have been put to use for the present study.

Soda glass was chosen as the host for the ion-exchange of silver. The composition (weight percent) of the glass used in the present study is Si-21.49%, Na-7.1%, Ca-5.78%, Mg-0.34% and Al-0.15%. Silver doped glass samples were prepared by immersing the preheated soda glass slides for 2 min in a molten salt bath of $AgNO_3$ and $NaNO_3$ (1:4 weight ratio) mixture at 320 $^oC$. These samples were postannealed for 1 hr in vacuum at a pressure lower than $1x10^{-6}$ mbar at various temperatures and subsequently cooled in the vacuum. For room temperature photoluminescence measurements in these samples, vertically polarized, 488 nm line of $Ar^+$ laser (Coherent) with 60 mW power was utilized to excite the samples and the scattered light from the sample was dispersed using a double monochromator (Spex 14018) and detected using a cooled photomultiplier tube (FW ITT 130) operated in the photon counting mode. Scanning of the spectra and data acquisition was carried out using a microprocessor based data-acquisition-cum-control system. The spectra were recorded from 100 to 8100 $cm^{-1}$ at 20 $cm^{-1}$ step with 10 s integration time. X-ray photoelectron spectroscopy (XPS) studies were carried out in an ultrahigh vacuum chamber ($1.7x10^{-10}$ mbar) with an Al-$K_\alpha$ X-ray source (1486.6 eV) using a hemispherical analyzer. The photoelectron spectrometer work function was adjusted to get the Au $4f_{7/2}$ peak at 84.0 eV. XPS analysis reveals the silver-exchanged glass sample as silver monoxide (AgO) (fig.1(a)). The binding energy and full width at half maximum (FWHM) of the Ag $3d_{5/2}$ peak is 367.5 eV and 1.3 eV (table I), respectively. This value is in good agreement with the published data [12,15]. In order to confirm further the chemical state of silver in the as-exchanged sample, the modified Auger parameter was found out analytically taking into account Ag $3d_{5/2}$ photoelectron peak and Ag $M_4N_{45}N_{45}$ Auger transition [16]. The parameter was estimated to be 724.9 eV in this sample and the value agrees quite well with that of AgO (724.8 eV) reported in the literature [17]. For comparison, the parameter was calculated to be 726.2 eV for bulk silver.

Room temperature optical absorption spectra were recorded using a Shimadzu dual-beam spectrophotometer (PC 3101). A plain glass slide was kept on the reference beam during the collection of absorption spectra. Figure 2 displays the optical absorption spectra for the samples before and after annealing at various temperatures. Absorption peaks observed in the silver-exchanged glass sample (see fig. 2(a)) are assigned to the exciton absorptions in AgO. This finding has not been reported earlier, to best of our knowledge. Photo excitation of electrons induces selective transitions from the valence band of AgO to higher energy states leaving holes behind and the excited electrons occupy different exciton energy levels close to the bottom of the conduction band of AgO. Formation of excitons in AgO is observed at room temperature. The exciton absorption peaks are at 2.15, 2.28 eV (fig. 2(a)). Apart from lifetime broadening, quasi-continuous exciton energy bands below the conduction band might be the possible reason of absorption widths at room temperature in this material. On annealing at higher temperatures (380 to 600 $^o$C), no significant feature between 2 to 2.6 eV is seen in the absorption spectra except the known excitation energy dependence (see fig. 2(b)-2(e)). Inset figure displays the surface-plasmon-resonance absorption at 3 eV due to the silver nanoparticles in the soda glass postannealed at various temperatures. During high temperature annealing, silver ions are released in the matrix due to the thermal decomposition of oxides of silver. Observed increase in the absorption intensity with the increase of annealing temperature (see inset figure) is due to the increase in volume fraction of silver nanoparticles in the glass. This was earlier verified with the help of Rutherford backscattering measurements [18].

Photoluminescence spectra of the silver-exchanged soda glass sample and after postannealing the samples at different temperatures are displayed in Fig. 3 along with the emission from the soda glass substrate. A dominant peak at 2.15 eV (577 nm) along with the other less intense peak at 2.28 eV (545 nm) are seen (see fig. 4(a)) for the as-exchanged soda glass sample. These peaks are assigned to the excitonic photoemissions in AgO. Photo excited (by 488 nm laser line) electrons undergo non-radiative thermal scattering processes before forming bound excitons (e-h pair) which are confined in space by Coulomb force and have enhanced probability for radiative recombination. Drastic changes in PL intensity are observed on postannealing the silver-exchanged glass samples at various temperatures (see fig. 3). PL intensity reaches the maximum due to postannealing at 450 $^o$C and falls to the minimum value after 600 $^o$C annealing. Broad photoemission bands centered around 1.95 eV (637 nm) and 2.23 eV (557 nm) are displayed for the samples postannealed at 380 $^o$C (fig. 4(b)) and 450 $^o$C (fig. 4(c)). While the emissions (red bands) lower than 2 eV are attributed to the presence of small silver nanoparticles [19], the emissions at 2.23 eV are ascribed to the band-to-band radiative transition in $Ag_2O$. Photoemission energy at 2.23 eV agrees extremely well with the optical band gap of $Ag_2O$ (2.25 eV) [19]. With the increase of silver concentration (due to Ag-O bond breaking) during annealing, it is natural to form a thermodynamically stable structure of $Ag_2O$ [20] and more evidences come from the present XPS data (table I). This was again investigated by calculating the modified Auger parameter for the postannealed samples following the same principle [16]. The parameter value was estimated to be 725.6 eV in these samples. The value is significantly higher than that of AgO and is attributed to the presence of $Ag_2O$ phase in these samples. XPS analysis (see figure 1 & table I) reveals that $Ag_2O$ phase was dominant near the surface of the postannealed samples and the presence of metallic silver was reflected in the peak broadening of Ag $3d_{5/2}$ photoelectrons.

Silver monoxide (AgO) formed in the as-exchanged sample is chemically unstable and could decompose into $Ag_2O$ and Ag when heated in high vacuum. However, thermochemical transformation due to progressive structural rearrangements of silver and oxygen during annealing the silver exchanged glass in a neutral atmosphere is not fully understood. XPS results particularly demonstrate the presence of Ag and $Ag_2O$ for the samples postannealed at higher temperatures (see fig. 1(b)-1(d)). XPS data indicate the chemical state of the surface (limited to few tens of angstrom) composition of the material and that has not really changed in cases of samples postannealed at 450 to 600 $^o$C (see table I). In the present case, depending on annealing temperature, the average size of silver nanoparticles varies in the range of 5 to 10 nm [18]. In this size range, screening of the core level holes created by the emission of photoelectrons (3d electrons) is possibly comparable to that of bulk silver and hence, no change in kinetic energy or, binding energy is expected with the increase of size of silver nanoparticles [21]. Thus, we establish that the observed changes in binding energies were as a result of oxidation of silver only.

Rapid rise of the PL intensity till 450 $^o$C (fig. 3) could be primarily due to the increase of volume fractions of $Ag_2O$ in the bulk of the glass. Thermochemical transformation of AgO into $Ag_2O$ might have occurred due to progressive structural rearrangements of Ag and O bonding during the thermal processing in this temperature range. Further increase of annealing temperature leads to the rapid growth of silver nanoparticles (as a result of thermal decomposition of $Ag_2O$) within a depth scale of about 100 nm from the glass surface [18] and that might have resulted in the quenching of PL intensity observed for the samples postannealed at 550 and 600 $^o$C. To explain the quenching phenomena further, we have carried out XPS measurements on a specially prepared sample: the sample (550 $^o$C annealed) was mechanically polished using 0.25 micron diamond paste for 5 min to remove a few layers from the surface of the material. This was performed to avoid the sputtering effects in XPS system for obtaining depth information. Interestingly, the recorded XPS spectra (fig.

1(d)) of this sample show significant changes in the peak profiles. Most markedly, the estimated ratio of $Ag_2O/Ag$ amount in the polished sample has reduced drastically to 0.6 from 2.7 when compared to the most luminescent sample (450 $^o$C annealed, fig. 3) in the present study. Following the same line of argument, we explain successive quenching of the PL intensity for the 600 $^o$C postannealed sample where silver nanoparticles had grown to an average size of 10 nm (HRTEM measurements, see Ref. 18). Here we would like to restate that all the experimental results discussed in this article evidently explained the quenching of photoluminescence intensity with the thermal growth of the silver nanoparticles in the soda glass matrix.

In conclusion, we have been able to explain the origin of photoluminescence and optical absorption carried out at room temperature in the silver-exchanged soda glass samples. A possible correlation of the growth of the silver nanoparticles and drastic changes in the photoluminescence intensity due to thermal annealing of the silver doped glass samples in a neutral atmosphere has been explained and demonstrated in the present report. The insights provided here might help in general to gain better perception about the novelty of the science of nanoscale materials.

The authors gratefully acknowledge M. Premila of Materials Science Division for her help in recording the optical absorption spectra.

**Table I:** Binding energy (B.E.) and full width at half maximum (FWHM) of Ag $3d_{5/2}$ photoelectron peaks at different stages of sample treatment.

| sample treatment | B.E. (FWHM) of Ag $3d_{5/2}$ (eV) | deconvoluted B. E. (FWHM) (eV) | remarks | $Ag_2O$/Ag ratio |
|---|---|---|---|---|
| as-exchanged | 367.5 (1.3) | - | AgO | - |
| 450 C annealed | 367.9 (1.6) | 367.8 (1.3) 368.4 (1.2) | $Ag_2O$ Ag | 2.7 |
| 600 C annealed | 367.8 (1.6) | 367.9 (1.3) 368.4 (1.3) | $Ag_2O$ Ag | 2.4 |
| 550 C annealed and polished | 368.3 (1.9) | 367.9 (1.3) 368.5 (1.5) | $Ag_2O$ Ag | 0.6 |

**Figure captions**

**FIG. 1.** Deconvoluted XPS spectra of Ag $3d_{5/2}$ photoelectrons from silver in soda glass samples, **(a)** as-exchanged, postannealed at **(b)** 450 °C, **(c)** 600 °C, and **(d)** polished sample after annealing at 550 °C. Low and high energy peaks in the XPS spectra correspond to $Ag_2O$ and Ag, respectively (for details, see Table I).

**FIG. 2. (Top) (a)** Formation of excitons in AgO is clearly seen in the optical absorption spectra for the silver-exchanged soda glass sample; **(Below)** Optical absorption spectra in the silver-exchanged soda glass samples due to postannealing at **(b)** 600 °C, **(c)** 500 °C, **(d)** 450 °C and **(e)** 380 °C. Inset figure shows the increase of SPR absorption at 3 eV due to the thermal growth of silver nanoparticles in the soda glass sample.

**FIG. 3. (a)** Room-temperature photoluminescence spectra from the silver-exchanged soda glass sample and from the postannealed samples at **(b)** 380 °C, **(c)** 450 °C, **(d)** 550 °C, **(e)** 600 °C along with the glass substrate.

**FIG. 4.** Experimental photoluminescence spectra (as seen in figure **3**) are further shown here as a sum of two Gaussian spectral functions. In the figure, symbols are the data and dashed lines are the fitted functions to the data.

FIG.1

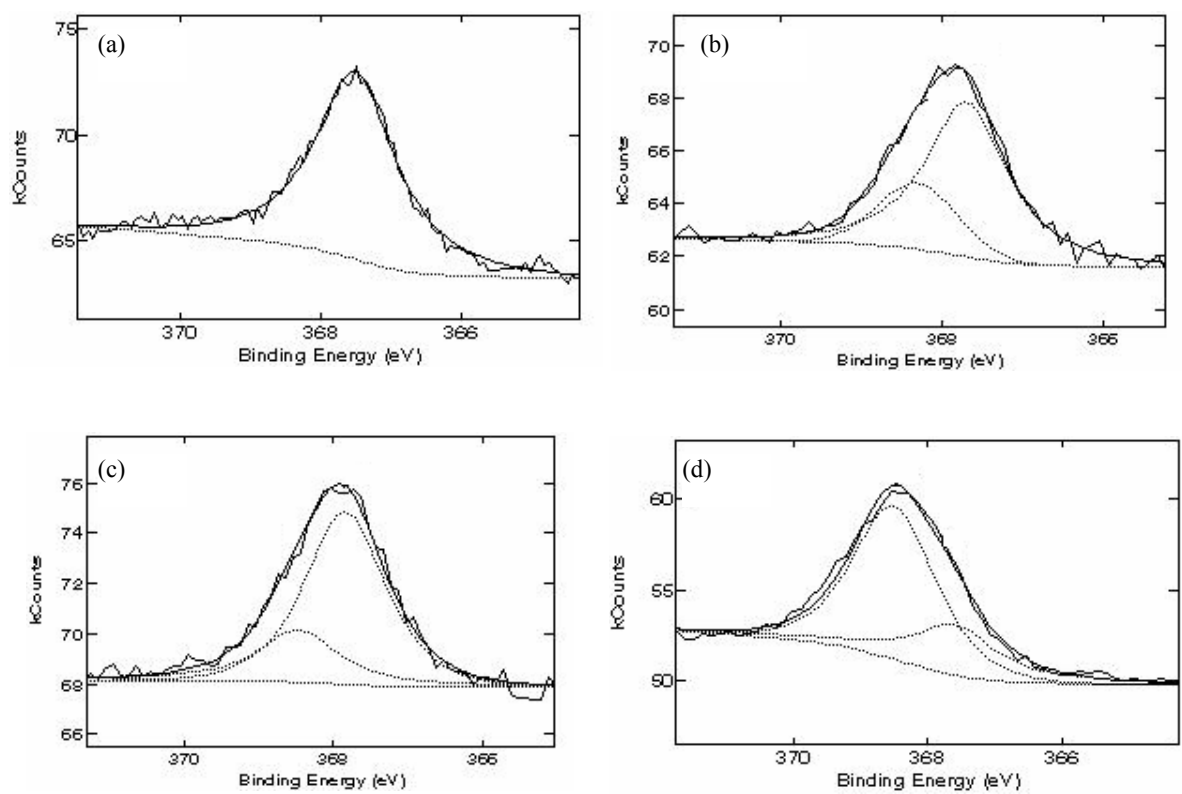

FIG.2

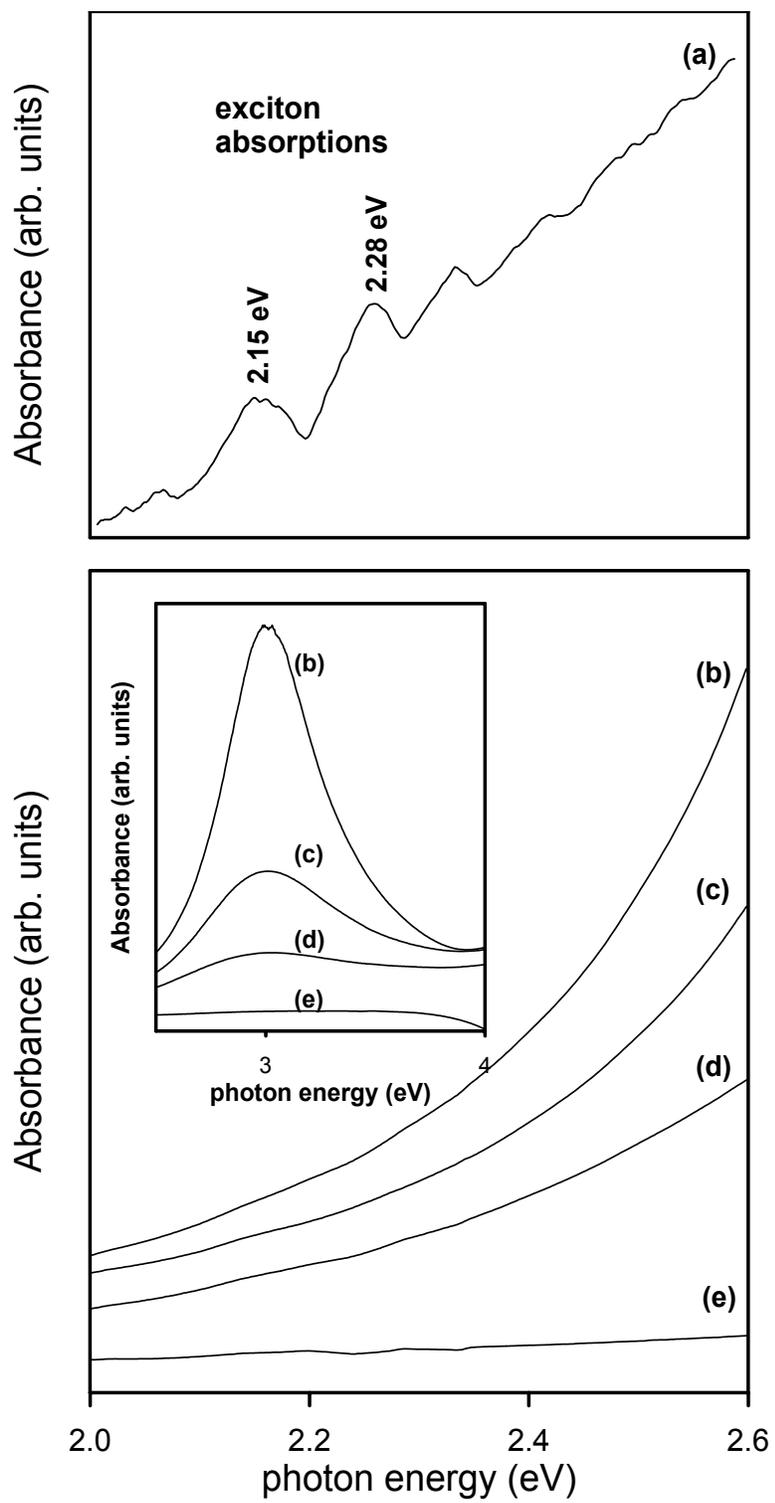

FIG.3

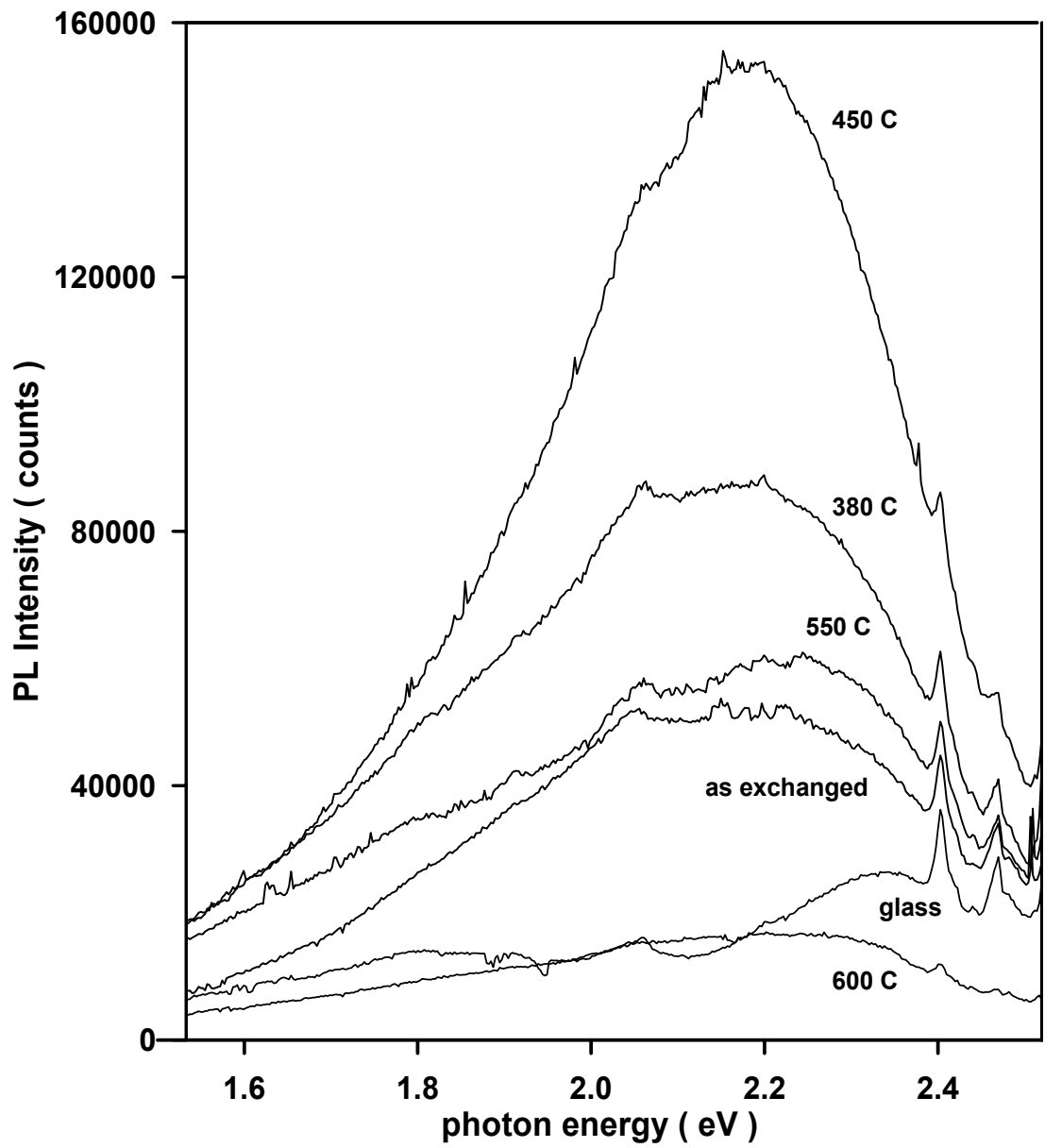

FIG.4

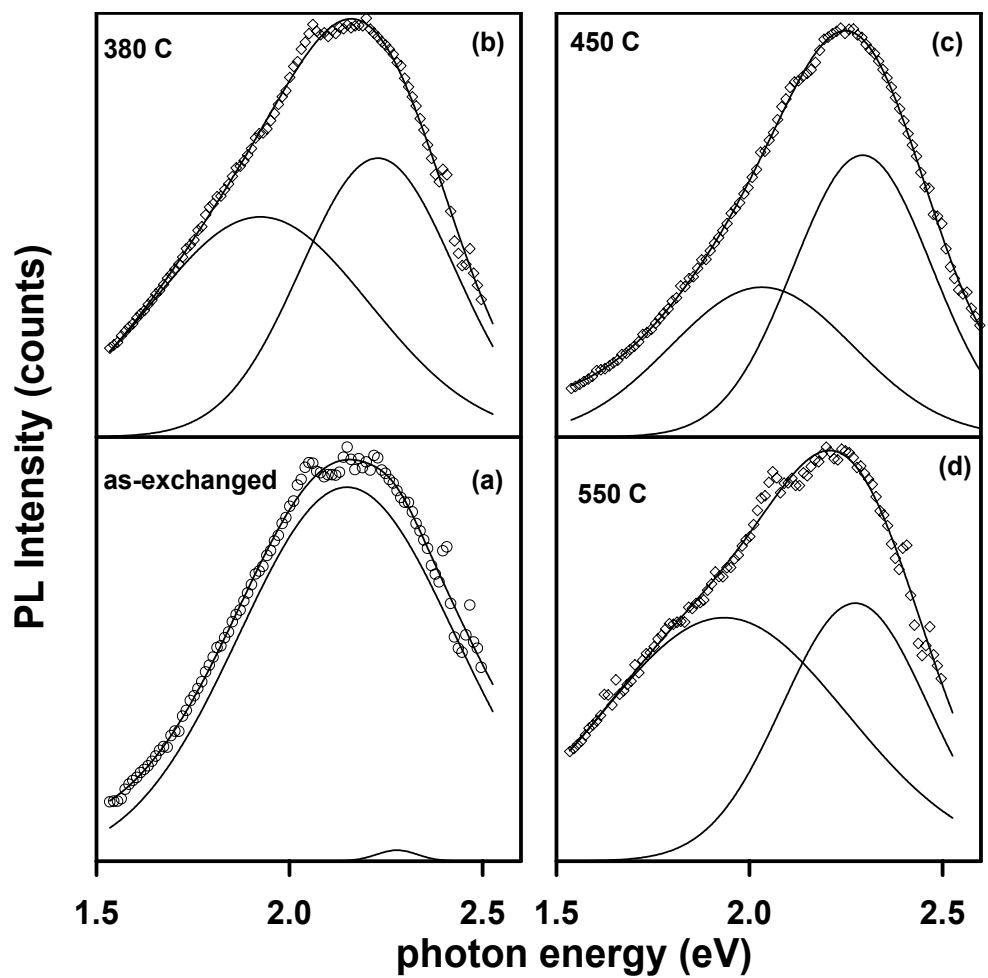